\documentclass[12pt]{amsart}
\usepackage{amsmath,amsfonts,amssymb} \usepackage{color}
\usepackage[all,cmtip]{xy}
\usepackage{graphicx}

\newcommand{\corr}[1]{\langle {#1} \rangle}

\newcommand{\Corr}[1]{\biggl\langle {#1} \biggr\rangle}

 \newcommand{\bP}{\mathbb{P}}  
   
 \newcommand{\cE}{\mathcal{E}}

   \newcommand{\cS}{\mathcal{S}}
 
 \newcommand{\cV}{\mathcal{V}}

 \newcommand{\bC}{\mathbb{C}}
 \newcommand{\pd}{\partial}
\newcommand{\Mbar}{\overline{\mathcal M}}

  \DeclareMathOperator{\res}{res}  
    \DeclareMathOperator{\Aut}{Aut}
\DeclareMathOperator{\tr}{tr} 

\newcommand{\be}{\begin{equation}}
\newcommand{\ee}{\end{equation}}
\newcommand{\bea}{\begin{eqnarray}}
\newcommand{\eea}{\end{eqnarray}}
\newcommand{\ben}{\begin{eqnarray*}}
\newcommand{\een}{\end{eqnarray*}}

\newcommand{\half}{\frac{1}{2}}

\newtheorem{cor}{Corollary}[section]

 \newtheorem{prop}[cor]{Proposition}
 \newtheorem{thm}[cor]{Theorem}
\theoremstyle{remark}

 \newtheorem{rmk}[cor]{Remark}

\definecolor{A}{rgb}{.75,1,.75}

\definecolor{green}{rgb}{0,1,0}
\definecolor{yellow}{rgb}{1,1,0}
\definecolor{orange}{rgb}{1,.7,0}
\definecolor{red}{rgb}{1,0,0}
\definecolor{white}{rgb}{1,1,1}

 \makeindex
\begin{document}
\title
[Virasoro Constraints as Residues]
{Virasoro Constraints of Curves as Residues}

\author{Jian Zhou}
\address{Department of Mathematical Sciences\\Tsinghua University\\Beijng, 100084, China}
\email{jianzhou@mail.tsinghua.edu.cn}

\begin{abstract}
Inspired by Eynard-Orantin topological recursions,
we reformulate the Virasoro constraints for curves
as residues of multilinear differentials.
As applications they can be used
to compute the $n$-point functions  
of Gromov-Witten invariants of curves.
\end{abstract}

\maketitle

\section{Introduction}
\label{sec:Introduction}

This is a sequel to an earlier work \cite{Zhou}
in which we reformulate the DVV Virarsoro constraints \cite{DVV}
for the Witten-Kontsevich tau-function \cite{Wit, Kon}
as Eynard-Orantin topological recursions on the Airy curve.
The topological recursions in Eynard-Orantin formalism are expressed
as residues of multilinear differentials on some spectral curve.
In this formalism one starts with a spectral curve and some initial data
and writes down the recursion relations in terms of residues,
and so the spectral curve has to be specified beforehand in order to write down the recursion relations.
It has been shown by Eynard and Orantin \cite{Eyn-Ora} that
if one applies the Eynard-Orantin formalism to the Airy curve
\begin{align}
x(z) & = \half z^2, & y(z) & = z, & B(z_1,z_2) & = \frac{dz_1dz_2}{(z_1-z_2)^2},
\end{align}
then the $n$-point functions are the $n$-point functions of the Witten-Kontsevich tau-function.
They use the Kontsevich model \cite{Kon} to achieve this.
An alternative way is to explicitly write down the Eynard-Orantin topological
recursions in this case in terms of the correlators,
and then match with the well-known Virasoro constraints for intersection
numbers on $\Mbar_{g,n}$ derived by Dijkgraaf-Verlinde-Verlinde \cite{DVV}.

The approach in \cite{Zhou} is different.
We start with the Virasoro constraints satisfied by the Witten-Kontsevich tau-function
and rewrite them in terms of residues after multiplying with some kernel function.
The kernel function is interpreted in terms of the genus zero two-point function
of the Gromov-Witten theory of a point,
while the Airy is given by the genus zero one-point function.
Since $\Mbar_{0,1}$ and $\Mbar_{0,2}$ do not exists,
one has to define the one-point function and the two-point function in genus zero
using the formula for $n$-point correlators in genus zero for $n \geq 3$.

Later we interpret the above two different approaches as
the reconstruction approach and the emergent approach respectively.
In general, in a reconstruction approach,
one starts with some building blocks and some fundamental laws
and reconstruct the whole theory;
in an emergent approach one starts with a given theory and uncover the building
blocks and the fundamental laws.

By emergent geometry of a Gromov-Witten type theory,
we mean the study of geometric structures that emerge from
the general properties of the theory.
Virasoro constraints and integrable hierarchies  
are the very general properties of Gromov-Witten theory
that are pursued for any given symplectic or projective manifold.
The emergent approach starts with one of these general properties
and seek for suitable geometric reformulations.

In this paper we will focus on the Virasoro constraints of a curve
proved by Okounkov and Pandharipande \cite{Oko-Pan-Vir}.
We show that by exactly the same approach as in \cite{Zhou}
one can reformulate their strengthened  Virasoro constraints
in terms of residue computations for $n$-point functions.

As applications one can compute the $n$-point functions of Gromov-Witten invariants
of curves.
The Virasoro constraints reduce the computations of the Gromov-Witten invariants
of a curves to its stationary sector.
The disconnected stationary GW invariants of a curves has been shown by Okounkov
and Pandharipande \cite{Oko-Pan-GW/H} using GW/Hurwitz correspondence
to be given by summations over partitions.
As consequences,
operator formulas for the disconnected  $n$-point functions for $\bP^1$
and elliptic curves are obtained.
In particular,
they give a method to compute the disconnected $n$-point function
for $\bP^1$ and they show that
the disconnected  $n$-point functions for elliptic curves
are given by the Bloch-Okounkov character formula \cite{Blo-Oko}.
The $n$-point functions of connected stationary GW invariants of $\bP^1$
are conjectured by Norbury and Scott \cite{Nor-Sco} to
satisfy the Eynard-Orantin topological recursion for
\begin{align}
x(z) & = z+ \frac{1}{z}, & y(z) & =\ln z, & B(z_1,z_2) & = \frac{dz_1dz_2}{(z_1-z_2)^2}.
\end{align}
This conjecture has been proved by Dunin-Barkowski {\em et al}
\cite{D-B-O-S-S}
and extended to the case of equivariant GW invariants by Fang, Liu and Zong \cite{Fan-Liu-Zon}.
More recently,
Dubrovin and Yang \cite{Dub-Yan} conjecture a formula for the $n$-point functions of
stationary GW invariants of the projective line.
This formula has been proved by Marchal \cite{Mar} and Dubrovin-Yang-Zagier \cite{Dub-Yan-Zag}
by different approaches.
By combining with the method in this paper,
one can then get formulas for $n$-point functions of
general GW invariants of the projective line.
The case of the projective line is particularly interesting 
because all $n$-point functions can now be computed using residues.
See also Borot-Norbury \cite{Bor-Nor}.

The rest of this note is arranged as follows.
In \S \ref{sec:Virasoro} we recall the strengthened Virasoro
constraints for algebraic curves.
We reformulate the Virasoro constraints and its strengthening
in \S \ref{sec:Vir} and \S \ref{sec:Str} respectively.
We recall in \S \ref{sec:P1} some methods to compute the $n$-point functions
of stationary GW invariants of $\bP^1$. They can be combined with our results. 
In \S \ref{sec:Conclusions} we present some concluding remarks
on some directions for further investigations.

\section{Strengthened Virasoro Constraints for Algebraic Curves}
\label{sec:Virasoro}

We now recall the Viroasoro constraints
and their strengthening
for algebraic curves proved by Okounkov and Pandharipande \cite{Oko-Pan-Vir}.
We will closely follow their notations with slight modifications.

\subsection{Virasoro constraints for algebraic curves}

Let $X$ be a nonsingular algebraic curve of genus $g$.
Let
\ben
&P \\
& \alpha_1, \dots, \alpha_g, \beta_1, \dots, \beta_g \\
& Q
\een
be a basis of $H^*(X, \bC)$ with the following properties:
\begin{itemize}
\item[(i)] the class $P=1 \in H^0(X, \bC)$ is the identity,
\item[(ii)] the classes $\alpha_i \in H^{1,0}(X, \bC)$ and $\beta_j \in H^{0,1}(X, \bC)$ determine a symplectic
basis of $H^1(X, \bC)$, i.e.,
\ben
\int_X \alpha_i \cup \beta_j = \delta_{ij},\een
\item[(iii)] The class $Q = \omega \in H^2(X, \bC)$ is the Poinca\'e dual of the point.
\end{itemize}
Introduce four sets of variables corresponding to the descendents of
these classes respectively:
\ben
&t^P_0, t^P_1, t^P_2, \dots, \\
& s^i_0, s^i_1, s^i_2, \dots, \bar{s}^j_0, \bar{s}^j_1, \bar{s}^j_2,\dots, \\
& t^Q_0, t^Q_1, t^Q_2, \dots,
\een
where $s^i_k$ and $\bar{s}^j_l$ are odd variables that supercommute.
Let $\xi$ denote the formal sum:
\be
\xi = \sum_{k\geq0}
t^P_k\tau_k(P) +\sum^g_{i=1} \sum_{k\geq0}\big(
s^i_k\tau_k(\alpha_i) + \bar{s}^i_k\tau_k(\beta_i)\big) +
\sum_{k\geq0} t^Q_k\tau_k(Q).
\ee
The free energy $F_X$ is the generating series of Gromov-Witten invariants of $X$:
\be
F_X = \sum_{n\geq 0}
\frac{1}{n!} \corr{\xi^n}_X,
\ee
the partition function $Z_X$ is
\be
Z_X = \exp F_X.
\ee

The Virasoro operators for the GW theory of $X$ are defined by:
\ben
L_{-1} &=& -\frac{\pd}{\pd t^P_0} + \sum_{n=0}^\infty
\biggl( t^P_{n+1}\frac{\pd}{\pd t^P_n}
+ \sum^g_{i=1} \biggl( s^i_{n+1}\frac{\pd}{\pd s^i_n}
+ \bar{s}^i_{n+1} \frac{\pd}{\pd \bar{s}^i_n} \biggr)
+ t^Q_{n+1}\frac{\pd}{\pd t^Q_n}\biggr) \\
& + & t^P_0t^Q_0 + \sum_{i=1}^g  s^i_0\bar{s}^i_0,
\een

\ben
L_0 & = & - \frac{\pd}{\pd t^P_1}
-\chi(X)\frac{\pd}{\pd t^Q_0} \\
& + & \sum_{n\geq0}
\biggl(nt^P_n \frac{\pd}{\pd t^P_n}
+ \sum^g_{i=1} \biggl((n + 1)s^i_n\frac{\pd}{\pd s^i_n} + n\bar{s}^i_n\frac{\pd}{\pd \bar{s}^i_n}\biggr)
+ (n + 1)t^Q_n \frac{\pd}{\pd t^Q_n} \biggr) \\
& + & \chi(X) \sum_{n\geq 0} t^P_{n+1} \frac{\pd}{\pd t^Q_n}
+\frac{\chi(X)}{2} t^P_0t^P_0,
\een
and for $k > 0$,
\ben
L_k &= & -(k+1)!\frac{\pd}{\pd t^P_{k+1}}
+ \sum_{n=1}^\infty \frac{(k+n)!}{(n-1)!}t^P_n \frac{\pd}{\pd t^P_{k+n}}
+ \sum_{n=0}^\infty \frac{(k+n+1)!}{n!} t^Q_n \frac{\pd}{\pd t^Q_{k+n}}\\
& + & \sum_{n=0}^\infty \sum^g_{i=1} \biggl(
\frac{(k+n+1)!}{n!} s^i_n\frac{\pd}{\pd s^i_{k+n}}
+ \frac{(n+k)!}{(n-1)!} \bar{s}^i_n
\frac{\pd}{\pd  \bar{s}^i_{k+n}} \biggr) \\
&-& \chi(X) (k+1)!\sum_{r=1}^{k+1}\frac{1}{r} \frac{\pd}{\pd t^Q_k}
+\chi(X) \sum_{n=1}^\infty \sum_{r=n}^{k+n}\frac{1}{r}
\frac{(k+n)!}{(n-1)!}t^P_n \frac{\pd}{\pd t^Q_{k+n-1}}\\
& + & \frac{\chi(X)}{2} \sum_{n=0}^{k-2} (n+1)!(k-n-1)!\frac{\pd}{\pd t^Q_n}\frac{\pd}{\pd t^Q_{k-n-2}}.
\een

\subsection{Strengthening of the standard Virasoro constraints}
Okounkov and Pandharipande also define two additional differential operators $D_k^i$ and $\bar{D}_k^i$ for
$k \geq -1$ by:
\ben
D^i_k & = & -(k + 1)!\frac{\pd}{\pd s^i_{k+1}}
+ \sum_{n\geq 0}
\frac{(k+n+1)!}{n!} t^P_n\frac{\pd}{\pd s^i_{k+n}} \\
& + & \sum_{n=0} \frac{(k+n+1)!}{n!} \bar{s}^i_n\frac{\pd}{\pd t^Q_{k+n}}, \\
\bar{D}^i_k & = & -(k + 1)!\frac{\pd}{\pd \bar{s}^i_{k+1}}
+ \sum_{n\geq 0} \frac{(k+n)!}{(n-1)!} t^P_n\frac{\pd}{\pd \bar{s}^i_{k+n}} \\
& - & \sum_{n\geq 0} \frac{(k+n+1)!}{n!} s^i_n\frac{\pd}{\pd t^Q_{k+n}}.
\een

The operators $L_k$, $D_k^i$, $\bar{D}_k^i$ annihilate the partition function $Z_X$,
and they satisfy the following commutation and anti-commutation relations:
\ben
&& [L_n, L_m] = (n - m)L_{n+m}, \\
&& [L_n, D_m^i ] = -(m + 1)D_{n +m}^i, \\
&& [L_n, \bar{D}_m^i ] = (n - m)\bar{D}_{n+m}^i, \\
&& \{D_n^i , D_m^j \} = \{D_n^i , \bar{D}_m^j \} = \{\bar{D}_n^i , \bar{D}_m^j \} = 0.
\een
If the operators $\{L_k\}_{k\geq -1}$ are identified with the Lie algebra of holomorphic vector field
\ben
\cV=\{-z^{k+1}\frac{\pd}{\pd z}\}_{k\geq 1},
\een
then, the operators $\{D_k^i \}_{k\geq -1}$ define a $\cV$-module
isomorphic to $\{-z^{k+1}\}_{k\geq-1}$
with the action defined by differentiation, and the operators $\{\bar{D}_k^i \}_{k\geq-1}$ define
a $\cV$-module isomorphic to the adjoint representation.

\section{Virasoro Constraints for Curves as Residues}
\label{sec:Vir}

\subsection{Virasoro constraints of curves in terms of correlators}

To illustrate the idea,
let us first set all the odd variables to zero for the moment.
Then the constraint $L_kZ_X=0$ can be rewritten in the following form:
\ben
&&k!\corr{\tau_k(P) \prod_{i=1}^n \tau_{a_i}(P) \cdot
\prod_{j=1}^m \tau_{b_j}(Q)} \\
& = & \sum_{i=1}^n \frac{(k+a_i-1)!}{(a_i-1)!} \corr{\tau_{k+a_i-1}(P)
\prod_{l\in [n]_i} \tau_{a_l}(P) \cdot \prod_{j=1}^m
\tau_{b_j}(Q)} \\
& + & \sum_{j=1}^m \frac{(k+b_j)!}{b_j!} \corr{\prod_{i=1}^n \tau_{a_i}(P) \cdot
\tau_{k+b_j-1}(Q) \prod_{l\in [m]_j} \tau_{b_l}(Q)} \\
& - & \chi(X) k! \sum_{r=1}^{k}\frac{1}{r}\corr{\tau_{k-1}(Q) \prod_{i=1}^n \tau_{a_i}(P) \cdot
\prod_{j=1}^m \tau_{b_j}(Q)}\\
& + & \chi(X)\sum_{i=1}^n \sum_{r=a_i}^{k+a_i-1} \frac{1}{r}
\frac{(k+a_i-1)!}{(a_i-1)!} \corr{\tau_{k+a_i-2}(Q) \prod_{l\in[n]_i} \tau_{a_l}(P) \cdot
\prod_{j=1}^m \tau_{b_j}(Q)} \\
& + & \frac{\chi(X)}{2} \sum_{n=0}^{k-3} (n+1)!(k-n-2)!
\corr{\tau_n(Q)\tau_{k-n-3}(Q)
 \prod_{i=1}^n \tau_{a_i}(P) \cdot
\prod_{j=1}^m \tau_{b_j}(Q)} \\
& + & \frac{\chi(X)}{2} \sum_{n=0}^{k-3} \sum_{I_1\coprod I_2=[n]} \coprod_{J_1\coprod J_2=[m]}
(n+1)!(k-n-2)!
\corr{\tau_n(Q) \tau_{a_{I_1}}(P)\tau_{b_{J_1}}(Q)} \\
&& \cdot \corr{\tau_{k-n-3}(Q)
\tau_{a_{I_2}}(P)\tau_{b_{J_2}}(Q)}.
\een
Here we have used the following notations:
$[n]$ denotes the set of indices $\{1, \dots, n\}$,
and for a subset $I \subset [n]$, $\tau_{a_I}(P) = \prod_{i\in I} \tau_{a_i}(P)$.
Similar notations will be used below.
The terms on the right-hand side can be interpreted as fusions and fissions of $\tau_k(P)$ respectively.

Multiply both sides by $\prod_{i=1}^n a_i! \prod_{j=1}^m (b_j+1)!$:
\ben
&&   \corr{\sigma_k(P) \sigma_{a_1}(P) \cdots \sigma_{a_n}(P)
\sigma_{b_1}(Q) \cdots \sigma_{b_m}(Q)} \\
& = & \sum_{i=1}^n a_i \corr{\sigma_{a_1}(P) \cdots \sigma_{k+a_i-1}(P) \cdots \sigma_{a_n}(P)
\sigma_{b_1}(Q) \cdots \sigma_{b_m}(Q)} \\
& + & \sum_{j=1}^m (b_j+1) \corr{\sigma_{a_1}(P) \cdots \sigma_{a_n}(Q)
\sigma_{b_1}(Q) \cdots \sigma_{k+b_j-1}(Q) \cdots \sigma_{b_m}(Q)} \\
& - & \chi(X)\sum_{r=1}^{k}\frac{1}{r} \corr{\sigma_{k-1}(Q) \sigma_{a_1}(P) \cdots \sigma_{a_n}(P)
\sigma_{b_1}(Q) \cdots \sigma_{b_m}(Q)} \\
& + & \chi(X)\sum_{i=1}^n \sum_{r=a_i}^{k+a_i-1} \frac{1}{r}
a_i \corr{\sigma_{a_1}(P) \cdots \sigma_{k+a_i-2}(Q) \cdots \sigma_{a_n}(P)
\sigma_{b_1}(Q) \cdots \sigma_{b_m}(Q)} \\
& + & \frac{\chi(X)}{2} \sum_{n=0}^{k-3}
\corr{\sigma_n(Q)\sigma_{k-n-3}(Q)\sigma_{a_1}(P) \cdots \sigma_{a_n}(P)
\sigma_{b_1}(Q) \cdots \sigma_{b_m}(Q)} \\
& + & \frac{\chi(X)}{2} \sum_{n=0}^{k-3} \sum_{I_1\coprod I_2=[n]} \coprod_{J_1\coprod J_2=[m]}
\corr{\sigma_n(Q) \sigma_{a_{I_1}}(P)\sigma_{b_{J_1}}(Q)} \\
&& \cdot \corr{\sigma_{k-n-3}(Q)\sigma_{a_{I_2}}(P)\sigma_{b_{J_2}}(Q)},
\een
where
\begin{align}
\sigma_a(P) & = a!\tau_a(P), & \sigma_b(Q) & = (b+1)!\tau_b(Q).
\end{align}

\subsection{Derivations of the recursion kernels}
We now rewrite the above recursion relations in terms of generating series:
\ben
&&  \sum_{k, a_i,b_j=0}^\infty
 \corr{\frac{\sigma_k(P)}{z_0^{k+1}} \frac{\sigma_{a_1}(P)}{z_1^{a_1+1}} \cdots \frac{\sigma_{a_n}(P)}{z_n^{a_n+1}}
\frac{\sigma_{b_1}(Q)}{w_1^{b_1+2}} \cdots \frac{\sigma_{b_m}(Q)}{w_m^{b_m+2}} } \\
& = &  \sum_{k, a_i,b_j=0}^\infty \sum_{i=1}^n \frac{a_iz_i^{k-1}}{z_0^{k+1}}
\corr{\frac{\sigma_{a_1}(P)}{z_1^{a_1+1}} \cdots \frac{\sigma_{k+a_i-1}(P)}{z_i^{k+a_i}}
\cdots \frac{\sigma_{a_n}(P)}{z_n^{a_n+1}}
\frac{\sigma_{b_1}(Q)}{w_1^{b_1+2}} \cdots \frac{\sigma_{b_m}(Q)}{w_m^{b_m+2}} } \\
& + &  \sum_{k, a_i,b_j=0}^\infty  \sum_{j=1}^m \frac{(b_j+1)w_j^{k-1}}{z_0^{k+1}} \corr{\frac{\sigma_{a_1}(P)}{z_1^{a_1+1}}
\cdots \frac{\sigma_{a_n}(Q)}{z_n^{a_n+1}}
\frac{\sigma_{b_1}(Q)}{w_1^{b_1+2}} \cdots \frac{\sigma_{k+b_j-1}(Q)}{w_j^{k+b_j+1}}
\cdots \frac{\sigma_{b_m}(Q)}{w_m^{b_m+2}} } \\
&  - &  \sum_{k, a_i,b_j=0}^\infty \chi(X)\sum_{r=1}^{k}\frac{1}{r}
 \corr{\frac{\sigma_{k-1}(Q)}{z_0^{k+1}} \frac{\sigma_{a_1}(P)}{z_1^{a_1+1}} \cdots \frac{\sigma_{a_n}(P)}{z_n^{a_n+1}}
\frac{\sigma_{b_1}(Q)}{w_1^{b_1+2}} \cdots \frac{\sigma_{b_m}(Q)}{w_m^{b_m+2}} } \\
& + & \chi(X)\sum_{i=1}^n \sum_{r=a_i}^{k+a_i-1} \frac{1}{r}
a_i \corr{\frac{\sigma_{a_1}(P)}{z_1^{a_1+1}} \cdots \frac{\sigma_{k+a_i-2}(Q)}{z_i^{k+a_i}}
 \cdots \sigma_{a_n}(P)
\sigma_{b_1}(Q) \cdots \sigma_{b_m}(Q)} \\
& + & \frac{\chi(X)}{2} \sum_{k=3}^\infty \sum_{n=0}^{k-3}
\corr{\frac{\sigma_n(Q)}{z_0^{n+2}} \frac{\sigma_{k-n-3}(Q)}{z_0^{k-n-1}} \prod_{i=1}^n \frac{\sigma_{a_i}(P)}{z_1^{a_i+1}} \cdot
\prod_{j=1}^m \frac{\sigma_{b_j}(Q)}{w_j^{b_j+2}}  } \\
& + & \frac{\chi(X)}{2} \sum_{n=0}^{k-3} \sum_{I_1\coprod I_2=[n]} \coprod_{J_1\coprod J_2=[m]}
\corr{\sigma_n(Q) \sigma_{a_{I_1}}(P)\sigma_{b_{J_1}}(Q)} \\
&& \cdot \corr{\sigma_{k-n-3}(Q)\sigma_{a_{I_2}}(P)\sigma_{b_{J_2}}(Q)}.
\een

If we introduce
\ben
&& W(u_1, \dots, u_n, x_1, \dots, x_m) \\
& = & \sum_{a_i,b_j=0}^\infty
 \corr{\frac{\sigma_{a_1}(P)}{u_1^{a_1+1}} \cdots \frac{\sigma_{a_n}(P)}{u_n^{a_n+1}}
\frac{\sigma_{b_1}(Q)}{x_1^{b_1+2}} \cdots \frac{\sigma_{b_m}(Q)}{x_m^{b_m+2}} }
du_1\cdots du_n dx_1\cdots dx_m,
\een
then
the above recursion relations can be rewritten in the following form:
\ben
&& W(u_0, u_1, \dots, u_n, x_1, \dots, x_m) \\
& = & \sum_{i=1}^n A(u;u_0,u_i) W(u_1, \dots, u_{i-1}, u, u_{i+1}, \dots, u_n, x_1, \dots, x_m) \\
& + & \sum_{j=1}^n B(x;u_0,x_i) W(u_1, \dots, u_n, x_1,\dots, x_{j-1}, x, x_{j+1},\dots, x_m) \\
& - & \chi(X) C(x;u_0) W(u_1, \dots, u_n, x, x_1, \dots, x_m) \\
& + & \sum_{i=1}^n \chi(X) D(x;u_0,u_i)W(u_1, \dots, u_{i-1},u_{i+1}, \dots, u_n, x, x_1, \dots, x_m) \\
& + & \frac{\chi(X)}{2}  E(x;u_0)
W(u_1, \dots, u_n, x, x, x_1, \dots, x_m) \\
& + & \sum_{I_1\coprod I_2=[n]} \coprod_{J_1\coprod J_2=[m]}\frac{\chi(X)}{2}  E(x;u_0)
W(u_{I_1}, x, x_{J_1})W(u_{I_2},x, x_{J_2}),
\een
where $A,B,C,D,E$ are operators to be specified below.

The operator $A(u;u_0,u_i)$  has the following effect:
\ben
&& \frac{1}{u^l}du \mapsto \sum_{k+a=l} \frac{au_i^{-a-1}}{u_0^{k+1}}du_i du_0.
\een
It describes a fusion of type $P + P \to P$, and
it can be realized by multiplying with the following differential form then take $\res_{u=0}$:
\ben
&& \sum_{l=0}^\infty u^{l-1}\sum_{k+a=l} \frac{au_i^{-a-1}}{u_0^{k+1}}du_i du_0
= \sum_{k,a=0}^\infty \frac{au^{k+a-1}u_i^{-a-1}}{u_0^{k+1}} du_i du_0 \\
& = & \sum_{a=0}^\infty au^{a-1}u_i^{-a-1} \cdot \sum_{k=0}^\infty \frac{u^k}{u_0^{k+1}} du_i du_0
= u_i^{-2} (1-\frac{u}{u_i})^{-2}  \cdot \frac{1}{u_0-u} du_i du_0 \\
& =  & \frac{du_i}{(u_i-u)^2} \cdot \frac{du_0}{u_0-u}.
\een

The operator $B(x;u_0,x_i)$ has the following effect:
\ben
&& \frac{1}{x^{l+1}}dx \mapsto \sum_{k+b=l} \frac{(b+1)x_j^{-b-2}}{u_0^{k+1}}d x_j du_0.
\een
It describe a fusion of type $P + Q \to Q$, and
it can be realized by multiplying with the following differential form then take $\res_{x=0}$:
\ben
&& \sum_{l=0}^\infty x^l \sum_{k+b=l} \frac{(b+1)x_j^{-b-2}}{u_0^{k+1}} dx_j du_0
= \sum_{k,b=0}^\infty \frac{(b+1)x_j^{-b-2}x^{k+b}}{u_0^{k+1}} dx_j du_0\\
& = &  \sum_{b=0}^\infty (b+1)x_j^{-b-2}x^{b} dx_j
\cdot  \sum_{k=0}^\infty \frac{x^{b}}{u_0^{k+1}}  du_0 \\
& = & \frac{dx_j}{(x_j-x)^2} \cdot \frac{du_0}{u_0-x}.
\een

The operator $C(x;u_0)$ has the following effect:
\ben
&& \frac{1}{x^{k+1}}dx \mapsto \frac{1}{u_0^{k+1}} \sum_{r=1}^k \frac{1}{r} du_0.
\een
It describes a conversion of type $P \to Q$.
It can be realized by multiplying with the following differential form then take $\res_{z=0}$:
\ben
&& \sum_{k=1}^\infty x^k \frac{1}{u_0^{k+1}} \sum_{r=1}^k \frac{1}{r} du_0 \\
&= & \sum_{r=1}^\infty \frac{1}{r} \sum_{k=r}^\infty \frac{x^k}{u_0^{k+1}} du_0
= \sum_{r=1}^\infty \frac{1}{r} \frac{x^r}{u_0^{r+1}} \frac{1}{1-\frac{x}{u_0}} du_0 \\
& = & -\log(1-\frac{x}{u_0}) \cdot \frac{du_0}{u_0-x}.
\een

The operator $D(x;u_0,u_i)$ has the following effect:
\ben
&& \frac{1}{x^{l}}dx \mapsto  \sum_{k+a=l,k, a\geq 1}
\frac{a}{u_0^{k+1}u_i^{a+1}} \sum_{r=a}^{k+a-1} \frac{1}{r}du_0du_i,
\een
It describes a fusion of type $P+P \to Q$.
It can be realized by multiplying with the following differential form then take $\res_{x=0}$:
\ben
&& \sum_{l=2}^\infty x^{l-1}
\sum_{k+a=l,a\geq 1} \frac{a}{u_0^{k+1}u_i^{a+1}} \sum_{r=a}^{k+a-1} \frac{1}{r}du_0du_i \\
& = & \sum_{r=1}^\infty \frac{1}{r}
\sum_{a=r}^\infty \frac{ax^{a-1}}{u_i^{a+1}}du_i
\sum_{k=0}^\infty \frac{x^{k}}{u_0^{k+1}}du_0 \\
& = & \sum_{r=1}^\infty \frac{1}{r} (\frac{x}{u_i})^{r-1}
\cdot\biggl( r- (r-1)\frac{x}{u_i}\biggr)\cdot \frac{du_i}{(u_i-x)^2}
\frac{1}{u_0}\frac{1}{1-\frac{x}{u_0}}du_0 \\
& =& \biggl(1-\log(1-\frac{x}{u_i})\biggr) \frac{du_i}{(u_i-x)^2}
\cdot \frac{du_0}{u_0-x}.
\een

The operator $E(x;u_0)$ has the following effect:
\ben
&& \frac{1}{x^{k+1}}(dx)^2 \mapsto \frac{du_0}{u_0^{k+1}}.
\een
It describes a fission of type $P \to Q + Q$.
It can be realized by multiplying with the following expression then take $\res_{u=0}$:
\ben
&& \sum_{k=0}^\infty \frac{x^k}{dx} \cdot \frac{du_0}{u_0^{k+1}}
= \frac{1}{dx} \cdot \frac{du_0}{u_0-x}.
\een

To summarize,
we have shown that:
\ben
&& W(u_0, u_{[n]}, x_{[m]}) \\
& = & \sum_{i=1}^n \res_{u=0} \biggl( \frac{du_idu}{(u_i-u)^2} \cdot
\frac{1}{du}\frac{du_0}{u_0-u}
  W(u, u_{[n]_i}, x_{[m]}) \biggr) \\
& + & \sum_{j=1}^n \res_{x=0} \biggl( \frac{dx_jdx}{(x_j-x)^2} \cdot
\frac{1}{dx} \frac{du_0}{u_0-x} \cdot W(u_{[n]}, x, x_{[m]_j})\biggr)  \\
& + & \res_{x=0} \biggl(\chi(X)\log(1-\frac{x}{u_0}) \cdot \frac{du_0}{u_0-x} \cdot
 W(u_{[n]}, x, x_{[m]})  \biggr) \\
& + & \res_{x=0} \biggl( \biggl(1-\log(1-\frac{x}{u_i})\biggr) \frac{du_idx}{(u_i-x)^2}
\cdot \frac{\chi(X)}{dx} \frac{du_0}{u_0-w}W(u_{[n]_i}, x, x_{[m]}) \biggr) \\
& + & \frac{1}{2}  \res_{x=0} \biggl(  \frac{\chi(X)}{dx} \cdot \frac{du_0}{u_0-x} \cdot
W(u_{[n]}, x, x, x_{[m]})  \biggr)\\
& + & \frac{1}{2}  \res_{w=0} \biggl(   \frac{\chi(X)}{dx} \cdot \frac{du_0}{u_0-x} \cdot
W(u_{I_1}, x, x_{J_1})W(u_{I_2},x, x_{J_2}) \biggr).
\een
Here we have used the following notations:
$[n]$ denotes the set of indices $\{1, \dots, n\}$, $u_{[n]}=u_1, \dots, u_n$;
$[n]_i =[n]-\{i\}$, $u_{[n]_i} = u_1, \dots, \widehat{u_i}, \dots, u_n$ ($u_i$ is deleted).

Now we can formulate the following

\begin{thm} \label{thm:Main}
The following recursion relations hold for nonsingular algebraic curves:
\ben
&& W(u_0, u_1, \dots, u_n, x_1, \dots, x_m) \\
& = & \sum_{i=1}^n \res_{u=0} \biggl(K(u_0,u) W_0(u_i,u)
  W(u, u_{[n]_i}, x_{[m]}) \biggr) \\
& + & \sum_{j=1}^n \res_{x=0} \biggl(K(u_0,x) W_0(x_j,x)
  W(u_{[n]}, x,x_{[m]_j})\biggr)  \\
& + & \res_{x=0} \biggl(\chi(X)L(u_0,x) \cdot
 W(u_{[n]}, x, x_{[m]})  \biggr) \\
& + & \sum_{i=1}^n \res_{x=0} \biggl(\chi(X)K(u_0,x) W_0(u_i,x)
W(u_{[n]_i}, x, x_{[m]}) \biggr) \\
& + & \frac{1}{2}  \res_{x=0} \biggl( \chi(X)K(u_0,x) \cdot
W(u_{[n]}, x, x, x_{[m]})  \biggr)\\
& + & \frac{1}{2}\sum_{I_1\coprod I_2=[n]} \coprod_{J_1\coprod J_2=[m]}
\res_{x=0} \biggl(  \chi(X)K(u_0,x) \cdot
W(u_{I_1}, x, x_{J_1})W(u_{I_2},x, x_{J_2}) \biggr),
\een
where
\ben
&& W_0(u_1, u_2) = \frac{du_1du_2}{(u_1-u_2)^2}, \\
&& W_0(x_1, x_2) = \frac{dx_1dx_2}{(x_1-x_2)^2}, \\
&& W_0(u, x)  = \biggl(1-\log\big(1-\frac{x}{u}\big) \biggr)dudx
\een
are genus zero two-point functions,
and
\ben
&& K(u_1,u_2) = \frac{1}{u_1-u_2} \frac{du_1}{du_2}, \\
&& K(u_1,x_2) = \frac{1}{u_1-x_2} \frac{du_1}{dx_2}, \\
&& L(u_1,x_2) = \log \big(1-\frac{x_2}{u_1}\big) \frac{du_1}{u_1-x_2}.
\een
\end{thm}

\begin{rmk}
$L(u_1,x_2)$ is related by $W_0(u_1,x_2)$ by differentiation:
\be
W_0(u_1,x_2) = -\frac{\pd L(u_1,x_2)}{\pd x_2}dx_2.
\ee
\end{rmk}

\subsection{Inclusion of the odd observables}

To understand the correlators with the insertions of odd observables,
for simplicity of presentation let us first consider
$$\corr{\tau_k(P) \tau_a(\alpha_i) \tau_b(\beta_j) }.$$
The Virasoro constraints $L_{k-1}Z_X =0$ give us:
\ben
&& k! \corr{\tau_k(P) \tau_a(\alpha_i) \tau_b(\beta_j) } \\
& = & \frac{(k+a)!}{a!} \corr{\tau_{k+a-1}(\alpha_i) \tau_b(\beta_j)}
+ \frac{(k+b-1)!}{(b-1)!} \corr{\tau_a(\alpha_i) \tau_{k+b-1}(\beta_j)}
\een
Multiplying both sides by $(a+1)!b!$ we get:
\ben
&& \corr{\sigma_k(P) \sigma_a(\alpha_i) \sigma_b(\beta_j) } \\
& = &  (a+1) \corr{\sigma_{k+a-1}(\alpha_i) \sigma_b(\beta_j)}
+ b \corr{\sigma_a(\alpha_i) \sigma_{k+b-1}(\beta_j)},
\een
where
\begin{align*}
\sigma_a(\alpha_i) & = (a+1)!\tau_a(\alpha_i), &
\sigma_b(\beta_j) & = b!\tau_b(\beta_j).
\end{align*}
In terms of the generating series:
\ben
&& \sum_{k,a,b\geq 0}
\corr{\frac{\sigma_k(P)}{u_0^{k+1}}
\frac{\sigma_a(\alpha_i)}{p_i^{a+2}} \frac{\sigma_b(\beta_j)}{\bar{p}_j^{b+1}} } \\
& = & \sum_{k,a,b\geq 0}
(a+1)\frac{p_i^{k-1}}{u_0^{k+1}} \corr{\frac{\sigma_{k+a-1}(\alpha_i)}{p_i^{k+a+1}}
\frac{\sigma_b(\beta_j)}{\bar{p}_j^{b+1}}} \\
& + & \sum_{k,a,b\geq 0} \frac{b\bar{p}_j^{k-1}}{u_0^{k+1}}
\corr{\frac{\sigma_a(\alpha_i)}{p_i^{a+2}} \frac{\sigma_{k+b-1}(\beta_j)}{\bar{p}_j^{k+b}}}.
\een
We reformulate it as follows:
\ben
W(u_0, p_i, \bar{p}_j)=F(q_i;u_0,p_i) W(q_i, \bar{p}_j)
+ G(\bar{q}_j;u_0, \bar{p}_j) W(p_i, \bar{q}_j),
\een
where the operators $F$ and $G$ will be specified below.

The operator $F(q_i;u_0,p_i)$ has the following effect:
\ben
\frac{dq_i}{q_i^{l+1}} \mapsto \sum_{k+a=l} (a+1) \frac{du_0}{u_0^{k+1}}\frac{dp_i}{p_i^{a+2}}.
\een
It describes the fusion of type $P + \alpha_i \to \alpha_i$.
It can be realized by multiplication with the following differential then take $\res_{v_i=0}$:
\ben
&& \sum_{l=0}^\infty q_i^{l} \sum_{k+a=l} (a+1) \frac{du_0}{u_0^{k+1}}\frac{dp_i}{p_i^{a+2}} \\
& = & \sum_{k=0}^\infty \frac{q_i^{k}}{u_0^{k+1}}dz_0 \sum_{a=0}^\infty (a+1) \frac{q_i^adp_i}{p_i^{a+2}}  \\
& = & \frac{du_0}{u_0-q_i} \frac{dp_i}{(p_i-q_i)^2} \\
& = & K(u_0, q_i) \cdot W_0(p_i, q_i),
\een
where
\ben
K(u_0, q_i) & = & \frac{du_0}{dq_i} \frac{1}{u_0-q_i}, \\
W_0(p_i, q_i) & = & \frac{dp_idq_i}{(p_i-q_i)^2}.
\een

The operator $G(\bar{q}_j;u_0,\bar{p}_j)$ has the following effect:
\ben
\frac{d\bar{q}_j}{\bar{q}_j^l} \mapsto
\sum_{k+b=l} \frac{bdu_0d\bar{p}_j}{u_0^{k+1}\bar{p}_j^{b+1}}.
\een
It describes the fusion of type $P + \beta_j \to \beta_j$.
It can be realized by multiplication with the following differential then take $\res_{\bar{q}_j=0}$:
\ben
\sum_{l=0}^\infty \bar{q}_j^{l-1} \sum_{k+b=l} \frac{bdz_0d\bar{p}_j}{u_0^{k+1}\bar{p}_j^{b+1}}
& = & \sum_{k=0}^\infty \frac{\bar{q}_j^k du_0}{u_0^{k+1}}
\sum_{b=0}^\infty \frac{b\bar{q}_j^{b-1}}{\bar{p}_j^{b+1}}
= \frac{du_0}{u_0-\bar{q}_j} \frac{d\bar{p}_j}{(\bar{p}_j-\bar{q}_j)^2} \\
& = & K(u_0, \bar{q}_j) W_0(\bar{p}_j, \bar{q}_j),
\een
where
\ben
&& K(u_0, \bar{q}_j) = \frac{du_0}{d\bar{q}_j} \frac{1}{u_0-\bar{q}_j}, \\
&& W_0(\bar{p}_i, \bar{q}_i) = \frac{d\bar{p}_id\bar{q}_i}{(\bar{p}_i-\bar{q}_i)^2}.
\een

Now we can generalize the recursion formula in Theorem \ref{thm:Main}
to the generating series
\ben
&& W(u, u_1, \dots, u_n, p_{i_1}^1, \dots, p_{i_k}^{k},\bar{p}_{j_1}^1, \dots, \bar{p}_{j_l}^l, x_1, \dots, x_m) \\
& = & \sum \biggl \langle \frac{\sigma_a(P)}{u^{a+1}}\frac{\sigma_{a_1}(P)}{u_1^{a_1+1}}
\cdots \frac{\sigma_{a_n}(P)}{u_n^{a_n+1}}
\frac{\sigma_{b_1}(\alpha_{i_1})}{(p_{i_1}^1)^{b_1+2}}  \cdots
\frac{\sigma_{b_k}(\alpha_{i_k})}{(p_{i_k}^k)^{b_k+2}} \\
&& \cdot \frac{\sigma_{c_1}(\beta_{j_1})}{(\bar{p}_{j_1}^1)^{c_1+1}}  \cdots
\frac{\sigma_{c_l}(\beta_{j_l})}{(\bar{p}_{j_l}^l)^{c_l+1}} \cdot
\frac{\sigma_{d_1}(Q)}{x_1^{d_1+2}}
\cdots \frac{\sigma_{d_m}(Q)}{x_m^{d_n+2}} \biggr\rangle.
\een
We omit the long formula which is awkward to present here.

\section{Strengthened Virasoro Constraints of Curves as Residues}
\label{sec:Str}

We now show how to reformulate the constraints given by operators
$D_k^i$ and $\bar{D}_k^i$ as residues.

\subsection{The case of $D_k^i$}

For simplicity of presentation,
we again consider a special case.
By the constraints $D_{k-1}Z_X = 0$ we get
\ben
k!\corr{\tau_k(\alpha_i)\tau_a(P)\tau_c(\beta_i)}
& = & \frac{(k+a)!}{a!} \corr{\tau_{k+a-1}(\alpha_i)\tau_c(\beta_i)} \\
& + & \frac{(k+c)!}{c!} \corr{\tau_a(P)\tau_{k+c-1}(Q)}.
\een
We multiply both sides by $(k+1) a!c!$ to get:
\ben
&& \corr{\sigma_k(\alpha_i)\sigma_a(P)\sigma_c(\beta^i)} \\
& = & (k+1) \corr{\sigma_{k+a-1}(\alpha_i)\sigma_c(\beta^i)}
+  (k+1)  \corr{\sigma_a(P)\sigma_{k+c-1}(Q)}.
\een
In terms of generating series we get:
\ben
&& \sum_{k,a,c \geq 0}
\corr{\frac{\sigma_k(\alpha_i)}{p_i^{k+2}} \frac{\sigma_a(P)}{u_1^{a+1}}
\frac{\sigma_c(\beta_i)}{\bar{p}_i^{c+1}}} \\
& = & \sum_{k,a,c \geq 0} \frac{(k+1) p_i^{a-1} }{u_1^{a+1}}
\corr{ \frac{\tau_{k+a-1}(\alpha_i)}{p_i^{k+a+1}} \frac{\sigma_c(\beta^i)}{\bar{p}_i^{c+1}} } \\
& + & \sum_{k,a,c \geq 0} \frac{ (k+1)\bar{p}_i^{k} }{p_i^{k+2} }
\corr{\frac{\sigma_a(P)}{u_1^{a+1}}\frac{\sigma_{k+c-1}(Q)}{\bar{p}_i^{k+c+1}}}.
\een
We rewrite it as follows:
\ben
W(p_i, u_1, \bar{p}_i)
=G(v_i;u_1, p_i)W(v_i, \bar{p}_i)
+ H(x;p_i,\bar{p}_i)W(u_1,x).
\een
Here the operator $G(q_i;u_1, p_i)$ has the following effect:
\ben
\frac{1}{q_i^{l+1}}dq_i \mapsto
\sum_{k+a=l} (k+1)\frac{dp_i}{p_i^{k+2}}\frac{du_1}{u_1^{a+1}}.
\een
It describes the fusion of type $\alpha_i + P \to \alpha_i$.
It can be realized by first multiplying with the following differential
then take $\res_{q_i=0}$:
\ben
&& \sum_{l=0}^\infty q_i^{l} \sum_{k+a=l} (k+1) \frac{1}{p_i^{k+2}}
\frac{1}{u_1^{a+1}} dp_idu_i
= \sum_{k=0}^\infty (k+1)\frac{v_i^{k}}{p_i^{k+2}}
\sum_{a=0}^\infty \frac{q_i^{a}}{u_1^{a+1}}dp_idu_i \\
& = & \frac{dp_i}{(p_i-q_i)^2} \cdot \frac{du_1}{u_1-q_i}
= K(u_1, q_i) \cdot W_0(p_i,q_i),
\een
where
\ben
&& W_0(p_i,q_i) = \frac{dp_idq_i}{(p_i-q_i)^2}, \\
&& K(u_1,q_i) = \frac{1}{u_1-q_i} \frac{du_1}{dq_i}.
\een
The operator $ H(x;p_i,\bar{p}_i)$ has the following effect:
\ben
\frac{1}{x^{l+1}}dx \mapsto
\sum_{k+c=l} (k+1) \frac{dp_i}{p_i^{k+1}}\frac{dq_i}{q_i^{c+1}}.
\een
It describes a fusion of type $\alpha_i + \beta_i \to Q$.
It can be realized by multiplying with the following differential
and taking $\res_{x=0}$:
\ben
&& \sum_{l=0}^\infty x^l \sum_{k+c=l}(k+1) \frac{dp_i}{p_i^{k+1}}\frac{d\bar{p}_i}{\bar{p}_i^{c+1}}
= \frac{dp_i}{(p_i-x)^2}\frac{d\bar{p}_i}{\bar{p}_i-x}
= K(\bar{p}_i,x) W_0(p_i,x),
\een
where
\ben
&& W_0(p_i,x) = \frac{dp_idx}{(p_i-x)^2}, \\
&& K(\bar{p}_i,x) = \frac{1}{\bar{p}_i-x} \frac{d\bar{p}_i}{dx}.
\een

\subsection{The case of $\bar{D}_k^i$}
We will again use an example to illustrate the idea.
By the constraints $\bar{D}_{k-1}^iZ_X = 0$ we have:
\ben
k!\corr{\tau_k(\beta_i)\tau_a(P)\tau_c(\alpha_i)}
& = & \frac{(k+a-1)!}{(a-1)!} \corr{\tau_{k+a-1}(\beta_i)\tau_c(\alpha_i)} \\
& - & \frac{(k+c)!}{c!} \corr{\tau_a(P)\tau_{k+c-1}(Q)}.
\een
We multiply both sides by $a!(c+1)!$ to get:
\ben
\corr{\sigma_k(\beta_i)\sigma_a(P)\sigma_c(\alpha_i)}
& = & a  \corr{\sigma_{k+a-1}(\beta_i)\sigma_c(\alpha_i)}
- (c+1) \corr{\sigma_a(P)\sigma_{k+c-1}(Q)},
\een
and so
\ben
&& \sum_{k,a,c \geq 0}
\corr{\frac{\sigma_k(\beta_i)}{\bar{p}_i^{k+1}}
\frac{\sigma_a(P)}{z_1^{a+1}}
\frac{\sigma_c(\alpha_i)}{p_i^{c+2}}} \\
& = & \sum_{k,a,c \geq 0} \frac{ap_i^{a-1} }{z_1^{a+1}}
\corr{ \frac{\tau_{k+a-1}(\beta_i)}{\bar{p}_i^{k+a}} \frac{\sigma_c(\alpha_i)}{p_i^{c+2}} } \\
& - & \sum_{k,a,c \geq 0} (c+1) \frac{p_i^{k-1}}{\bar{p}_i^{k+1}}
\corr{\frac{\sigma_a(P)}{z_1^{a+1}}\frac{\sigma_{k+c-1}(Q)}{p_i^{k+c+1}}}.
\een
So we need an operator such that
\ben
\frac{1}{\bar{q}_i^{l}}d\bar{q}_i \mapsto \sum_{k+a=l} a\frac{d\bar{p}_i}{\bar{p}_i^{k+1}}
\frac{du_1}{u_1^{a+1}},
\een
from this we get:
\ben
&& \sum_{l=0}^\infty \bar{q}_i^{l-1}
\sum_{k+a=l} \frac{1}{\bar{p}_i^{k+1}} \frac{a}{u_1^{a+1}} d\bar{p}_idu_1
= \sum_{k=0}^\infty \frac{\bar{q}_i^{k}}{p_i^{k+1}}
\sum_{a=0}^\infty \frac{a\bar{q}_i^{a-1}}{u_1^{a+1}}dp_idu_1 \\
& = & \frac{d\bar{p}_i}{\bar{p}_i-\bar{q}_i} \cdot
\frac{du_1}{(u_1-\bar{q}_i)^2}
= K(p_i,q_i) W_0(u_1, \bar{q}_i),
\een
where
\ben
&& W_0(u_1, \bar{q}_i) = \frac{du_1d\bar{q}_i}{(u_1-\bar{q}_i)^2}, \\
&& K(p_i,q_i) = \frac{1}{p_i-q_i} \frac{du_i}{dq_i}.
\een
We also need an operator such that:
\ben
\frac{1}{x^{l+1}}dx \mapsto \sum_{k+c=l}(c+1) \frac{d\bar{p}_i}{\bar{p}_i^{k+1}}
\frac{dp_i}{p_i^{c+2}}
\een
From this we get
\ben
&& \sum_{l=0}^\infty x^l \sum_{k+c=l} (c+1) \frac{d\bar{p}_i}{\bar{p}_i^{k+1}}
\frac{dp_i}{p_i^{c+2}}
= \frac{d\bar{p}_i}{\bar{p}_i-x}\frac{dp_i}{(p_i-x)^2},
\een
and so
\ben
&& W_0(p_i,x) =\frac{dp_idx}{(p_i-x)^2}, \\
&& K(\bar{p}_i, x) = \frac{d\bar{p}_i}{dx} \frac{1}{\bar{p}_i-x}.
\een

From these examples,
the reformulation for the general case can be easily deduced.

\section{Applications to $\bP^1$}

\label{sec:P1}

In this Section we will present the applications to the case of $\bP^1$.
We first recall the three different methods to compute the $n$-point
functions of stationary GW invariants of $\bP^1$ mentioned earlier,
then we note the explicit formulas obtained by these methods
to carry out the computations
for general $n$-point functions of GW invariants
of $\bP^1$ using our formula.

\subsection{The formula of Okounkov-Pandharipande}
We refer to the original paper by Okounkov-Pandharipande \cite{Oko-Pan-GW/H}
for definitions and notations.
They define the relative GW invariants of $\bP^1$
parameterized by two partitions $\mu$, $\nu$ of the same size:
\be
F_{\mu,\nu}^\bullet(z_1,\cdots, z_n) =
\sum_{k_1,\dots,k_n=-2}^\infty \Corr{\mu, \prod_{i=1}^n \tau_{k_i}(Q), \nu}
\prod_{i=1}^n z_i^{k_i+1},
\ee
then they show that they are given by expectation values of some operators
on the fermionic Fock space:
\be
F^\bullet_{\mu,\nu}(z_1,\cdots, z_n) = \Corr{\prod \alpha_{\mu_i}
\prod_{i=1}^n \cE_0(z_i) \prod \alpha_{-\nu_j}}.
\ee
Their formula for one-point function is
\be
\begin{split}
& \sum_{g=0}^\infty z^{2g} \corr{\mu, \tau_{2g-2+l(\mu)+l(\nu)}(Q), \nu}^\circ \\
= &\frac{1}{|\Aut(\mu)| |\Aut(\nu)|} \frac{\prod \cS(\mu_iz) \prod \cS(\nu_iz)}{\cS(z)},
\end{split}
\ee
where, by definition,
\be
\cS(z) = \frac{\sinh(z/2)}{z/2}
= \sum_{k=0}^\infty
\frac{z^{2k}}{2^{2k} (2k + 1)!}.
\ee
In particular, for $\mu = \nu = (1^d)$,
\be \label{eqn:P1-1-point}
\sum_{g=0}^\infty
z^{2g} \corr{\tau_{2g-2+2d}(Q)}^{\bP^1}_{g,1;d} = \frac{1}{(d!)^2} \cS(z)^{2d-1}.
\ee
They also give an algorithm to inductively compute the $n$-point functions.
An alternative way they give is to show that a certain partition function satisfies
the Toda equations,
and this are equivalent to the following recurrence relation for n-point functions.
For any partitions $\mu$ and $\nu$ of the same size, we have
\be
\begin{split}
&F^{\circ}_{\mu+1,\nu+1}(z_1,\dots, z_n) \\=&
\frac{1}{(m_1(\mu) + 1)(m_1(\nu) + 1)} \sum_{ \{(S_i,\mu^i,\nu^i)\}}
\prod_i \zeta(\Sigma_{S_i})^2 F^\circ_{\mu^i,\nu^i}(z_{S_i}) ,
\end{split}
\ee
where $\zeta(z) = 2\sinh(z/2)$,
and the summation is over all sets of triples
$\{(S_i, \mu^i, \nu^i)\}$,
such that $\{S_i\}$ is a partition of the set $[n]=\{1, \dots, n\}$ into nonempty disjoint
subsets:
\ben
&& [n] =  \coprod S_i , \quad S_i \neq \emptyset,
\een
similarly, $\{\mu^i\}$ and $\{\nu^i\}$ satisfy
\begin{align*}
\mu & = \coprod \mu^i , & \nu & = \coprod \nu^i , & |\mu^i| & = |\nu^i|,
\end{align*}
and where, by definition, $z_S = \{z_i\}_{i\in S}$ and $\Sigma_S = \sum_{i\in S} z_i$.

The following are some samples of computations using this formula:
\ben
F_{(1),(1)}^\circ(z_1,z_2) & = &
\zeta(z_1+z_2)^2F_{\emptyset,\emptyset}(z_1,z_2) \\
& + & \zeta(z_1)^2F^\circ_{\emptyset,\emptyset}(z_1) \cdot
\zeta(z_2)^2F^\circ_{\emptyset,\emptyset}(z_2) \\
& = & \zeta(z_1)^2\cdot \frac{1}{\zeta(z_1)} \cdot
\zeta(z_2)^2\cdot \frac{1}{\zeta(z_2)} \\
& = & \zeta(z_1)\zeta(z_2),
\een

\ben
F_{(1^2),(1^2)}^\circ(z_1,z_2)& = & \frac{1}{2\cdot 2}
\biggl[\zeta(z_1+z_2)^2 F^\circ_{(1),(1)}(z_1,z_2) \\
& + & \zeta(z_1)^2F^\circ_{\emptyset,\emptyset}(z_1) \cdot
\zeta(z_2)^2F^\circ_{(1),(1)}(z_2)  \\
& + & \zeta(z_1)^2F^\circ_{(1),(1)}(z_1) \cdot
\zeta(z_2)^2F^\circ_{\emptyset,\emptyset}(z_2)\biggr]\\
& = & \frac{1}{4} \biggl[\zeta(z_1+z_2)^2\zeta(z_1)\zeta(z_2) \\
& + & \zeta(z_1)^2\cdot \frac{1}{\zeta(z_1)} \cdot
\zeta(z_2)^2\cdot  \zeta(z_2) \\
& + & \zeta(z_1)^2\cdot  \zeta(z_1)  \cdot
\zeta(z_2)^2\cdot \frac{1}{\zeta(z_2)} \biggr] \\
& = & \frac{\zeta(z_1)\zeta(z_2)}{2!^2} [\zeta(z_1+z_2)^2
+ (\zeta(z_1)^2+\zeta(z_2)^2)].
\een
Here is another example:
\ben
F_{(1^3),(1^3)}^\circ(z_1,z_2) & = & \frac{\zeta(z_1)\zeta(z_2)}{3!^2} [\zeta(z_1+z_2)^4
+ \zeta(z_1+z_2)^2(\zeta(z_1)^2+\zeta(z_2)^2) \\
& + & (\zeta(z_1)^4 +   2!^2 \zeta(z_1)^2\zeta(z_2)^2+ \zeta(z_2)^4) ].
\een
In general,
\be \label{eqn:P1-2-point}
\begin{split}
& F_{(1^m),(1^m)}^\circ(z_1,z_2) \\
= & \frac{\zeta(z_1)\zeta(z_2)}{m!^2}
\sum_{k=0}^{m-1} \zeta(z_1+z_2)^{2m-2-2k}
\sum_{j=0}^k \binom{k}{j}^2
 \zeta(z_1)^{2j}\zeta(z_2)^{2k-2j}.
\end{split}
\ee

\subsection{One-Point function  by the formula of Okounkov-Pandharipande}
\label{sec:OP-One-Point}

For the one-point function by \eqref{eqn:P1-1-point}
one has
\ben
&& \sum_{g=0}^\infty
\lambda^{2g} \sum_{d=0}^\infty q^d
\corr{\tau_{2g-2+2d}(Q)}^{\bP^1}_{g,1;d}
\cdot (2g-1+2d)! x^{-2g-2d}\\
& = & \sum_{g=0}^\infty
\lambda^{2g} \sum_{d=0}^\infty q^d
\cdot (2g-1+2d)! x^{-2g-2d}
\res_{w=0} \frac{\cS(w)^{2d-1}}{w^{2g+1}d!^2}
\een
Now one can proceed in two different ways.
First one can take the summation over $d$ to get:
\ben
& &  \sum_{g=0}^\infty   \lambda^{2g}
\res_{w=0} \biggl( \frac{\cS(w)^{-1}}{z^{2g}w^{2g+1}}
\sum_{d=0}^\infty q^d
\cdot \frac{(2g-1+2d)!}{d!^2} x^{-2d}
\cS(w)^{2d} \biggr) \\
& = & \res_{w=0}
\biggl(- \frac{\cS(w)^{-1}}{w}
\ln \frac{1+\sqrt{1-4q\cS(w)^2/x^2}}{2} \biggr) \\
& + & \sum_{g=1}^\infty \lambda^{2g}
\res_{w=0} \biggl(\frac{\cS(w)^{-1}}{x^{2g}w^{2g+1}}
\frac{(2g-1)!}{(1-4 q\cS(w)^2/x^2)^{(4g-1)/2}}
 \sum_{j=0}^{g-1}
\frac{(2g-1)!(q\cS(w)^2/x^2)^j}{(2g-1-2j)!j!j!} \biggr) \\
& = &  -\biggl(\ln \frac{x+\sqrt{x^4-4q}}{2} - \ln x\biggr) \\
& + & \lambda^2
\frac{16q/x^2-1}{24z^2(1-4q/x^2)^{5/2}} \\
& + & \lambda^4\frac{18432(q/x^2)^3+8256(q/x^2)^2-94q/x^2+7}
{960x^4(1-4q/x^2)^{11/2}} \\
& + & \lambda^6\frac{1}{8064x^6(1-4q/x^2)^{17/2}}[-31+1180q/x^2+134886(q/x^2)^2 \\
& + & 5419360(q/x^2)^3+23229440(q/x^2)^4+13271040(q/x^2)^5 ]
+ \cdots.
\een
From this one can see that for $g=0$,
\be \label{eqn:P1-One-Point-Gen}
\sum_{d=1}^\infty q^d
\corr{\tau_{2d-2}(Q)}^{\bP^1}_{0,1;d}
\cdot (2d-1)! x^{-2d}
= \ln x -\ln \frac{x+\sqrt{x^2-4q}}{2},
\ee
and for $g\geq 1$,
$$\sum_{d=0}^\infty q^d
\corr{\tau_{2g-2+2d}(Q)}^{\bP^1}_{g,1;d}
\cdot (2g-1+2d)! x^{-2g-2d}$$
has the shape
\ben
\frac{1}{x^{2g+2}(1-4q/x^2)^{(6g-1)/2}}\sum_{j=0}^{2g-1} a_{g,j}(q/x^2)^j.
\een

According to \cite{OEIS},
the coefficients
$$\frac{n!}{(n-2j)!j!j!}$$
 are the sequence A089627.
They are the $\gamma$-vectors of n-dimensional type B associahedra.
This suggests some possibility of connection with the theory of cluster algebras.
They have the following generating series:
 \ben
&& \sum_{n=0}^\infty \sum_{j=0}^{[n/2]}\frac{n!}{(n-2j)!j!j!} x^ny^j
=\sum_{m=0}^\infty x^m \sum_{j=0}^\infty \frac{(m+2j)!}{m!j!^2}(x^2y)^j \\
& = & \sum_{j=0}^\infty \frac{(x^2y)^j}{j!^2} \sum_{m=0}^\infty \frac{(m+2j)!}{m!}x^m
= \sum_{j=0}^\infty \frac{(x^2y)^j}{j!^2} \frac{(2j)!}{(1-x)^{2j}} \\
& = & \frac{1}{\sqrt{1 -\frac{4x^2y}{(1-x)^2}}},
\een
therefore,
\ben
&& \sum_{g=0}^\infty \sum_{j=0}^{g-1}\frac{(2g-1)!}{(2g-1-2j)!j!j!} x^{2g-1} y^j
= \frac{1}{2}\frac{1}{\sqrt{1 -\frac{4x^2y}{(1-x)^2}}}
-\frac{1}{2}\frac{1}{\sqrt{1 -\frac{4x^2y}{(1+x)^2}}}.
\een
Unfortunately,
there is an extra factor $(2g-1)!$ that prevents us to use this summation to simplify
the above computations.

Secondly for $d > 0$ one can expand $\cS(w)^{2d-1}$ as follows:
\ben
\cS(w)^{2d-1}
& = & \frac{1}{w^{2d-1}} (e^{w/2}-e^{w/2})^{2d-1} \\
& = & \frac{1}{w^{2d-1}} \sum_{l=0}^{d-1}
\binom{2d-1}{l} (-1)^l (e^{(2d-1-2l)w/2}-e^{-(2d-1-2l)w/2}) \\
& = & \frac{1}{w^{2d-1}} \sum_{l=0}^{d-1}
\binom{2d-1}{l} (-1)^l
\sum_{j=0}^\infty \frac{(2d-1-2l)^{2j+1}}{2^{2j}(2j+1)!}w^{2j+1},
\een
and for $d=0$,
\ben
\cS(w)^{-1}
& = & \frac{w}{e^{w/2}-e^{w/2}}
= \sum_{m=0}^\infty \frac{B_{2m}}{(2m)!}(2^{1-2m}-1)w^{2m},
\een

\ben
&& \sum_{g=0}^\infty
\lambda^{2g} \sum_{d=0}^\infty q^d\corr{\tau_{2g-2+2d}(Q)}^{\bP^1}_{g,1;d}
\cdot (2g-1+2d)! z^{-2g-2d} \\
& = & \sum_{g=0}^\infty
\lambda^{2g} \cdot (2g-1)! x^{-2g} \frac{B_{2g}}{(2g)!}(2^{1-2g}-1) \\
& + & \sum_{j=0}^\infty \sum_{\substack{g+d=j+1\\g \geq 0, d\geq 1}}
\lambda^{2g} q^d(2g-1+2d)!x^{-2g-2d}\\
&& \cdot
\frac{1}{d!^2} \sum_{l=0}^{d-1}
\binom{2d-1}{l} (-1)^l  \frac{(2d-1-2l)^{2j+1}}{2^{2j}(2j+1)!} \\
& = & \sum_{g=0}^\infty
\lambda^{2g} \cdot x^{-2g} \frac{B_{2g}}{2g}(2^{1-2g}-1) \\
& + & \sum_{j=0}^\infty \frac{1}{4^jx^{2j+2}} \sum_{d=1}^\infty
\lambda^{2(j+1-d)}
\frac{q^d}{d!^2} \sum_{l=0}^{d-1}
\binom{2d-1}{l} (-1)^l (2d-1-2l)^{2j+1}.
\een

\subsection{Two-point function by the formula of Okounkov-Pandharipande}
\label{sec:OP-Two-Point}

By \eqref{eqn:P1-2-point} we get
\ben
&& \sum_{g=0}^\infty
\sum_{k_1,k_2 \geq 0} \corr{\tau_{k_1}(Q)\tau_{k_2}(Q)}_{g,2;d} z_1^{k_1+1}z_2^{k_2+1} \\
& = & \frac{\zeta(z_1)\zeta(z_2)}{d!^2}
\sum_{k=0}^{d-1} \zeta(z_1+z_2)^{2d-2-2k}
\sum_{j=0}^k \binom{k}{j}^2
 \zeta(z_1)^{2j}\zeta(z_2)^{2k-2j}.
\een

By the selection rule:
\be
(k_1 + 1) + (k_2+1) = 2g-2 + 2d + 2 = 2(g+d),
\ee
therefore,
\be \label{eqn:P1-Two-Point}
\begin{split}
& \sum_{d=0}^\infty q^d \sum_{g=0}^\infty \lambda^{2g}
\sum_{k_1,k_2 \geq 0} \corr{\tau_{k_1}(Q)\tau_{k_2}(Q)}_{g,2;d} z_1^{k_1+1}z_2^{k_2+1} \\
= & \sum_{d=0}^\infty q^d \lambda^{-2d} \sum_{g=0}^\infty
\sum_{k_1,k_2 \geq 0} \corr{\tau_{k_1}(Q)\tau_{k_2}(Q)}_{g,2;d}
(\lambda z_1)^{k_1+1}(\lambda z_2)^{k_2+1} \\
= & \sum_{d=0}^\infty q^d  \frac{\zeta(\lambda z_1)\zeta(\lambda z_2) }{d!^2\lambda^{2d}}
\sum_{k=0}^{d-1} \zeta(\lambda z_1+\lambda z_2)^{2d-2-2k} \\
& \cdot \sum_{j=0}^k \binom{k}{j}^2
 \zeta(\lambda z_1)^{2j}\zeta(\lambda z_2)^{2k-2j}.
\end{split}
\ee
If we consider the terms with $\lambda^0$,
we get:
\ben
&& \sum_{d=0}^\infty q^d
\sum_{k_1,k_2 \geq 0} \corr{\tau_{k_1}(Q)\tau_{k_2}(Q)}_{0,2;d} z_1^{k_1+1}z_2^{k_2+1} \\
& = & \sum_{d=0}^\infty q^d  \frac{z_1z_2}{d!^2}
\sum_{k=0}^{d-1} (z_1+z_2)^{2d-2-2k}
\sum_{j=0}^k \binom{k}{j}^2 z_1^{2j} z_2^{2k-2j} \\
& = & \sum_{d=0}^\infty q^d  \frac{z_1z_2}{d!^2}
\sum_{k=0}^{d-1} \sum_{l=0}^{2d-2-2k} \binom{2d-2-2k}{l}
z_1^lz_2^{2d-2-2k-l}
\sum_{j=0}^k \binom{k}{j}^2 z_1^{2j} z_2^{2k-2j} \\
& = & \sum_{d=0}^\infty    \frac{q^d}{d!^2}
\sum_{k=0}^{d-1} \sum_{l=0}^{2d-2-2k} \binom{2d-2-2k}{l}
\sum_{j=0}^k \binom{k}{j}^2 z_1^{2j+l+1} z_2^{2d-1-l-2j}.
\een
Because $k_1+k_2=2d-2$ is even,
we have two cases to consider.
Case 1. $k_1=2j+l=2n_1$, $k_2=2d-2-l-2j=2n_2$,
\ben
&& \corr{\tau_{2n_1}(Q)\tau_{2n_2}(Q)}_{0,2;n_1+n_2+1} \\
& = & \frac{q^{n_1+n_2+1}}{(n_1+n_2+1)!^2}
\sum_{k=0}^{n_1+n_2} \sum_{j=0}^k   \binom{2n_1+2n_2-2k}{2n_1-2j}
 \binom{k}{j}^2 .
\een
Case 2. $k_1=2j+l=2n_1-1$, $k_2=2d-2-l-2j=2n_2-1$,
\ben
&& \corr{\tau_{2n_1-1}(Q)\tau_{2n_2-1}(Q)}_{0,2;n_1+n_2} \\
& = & \frac{q^{n_1+n_2}}{(n_1+n_2)!^2}
\sum_{k=0}^{n_1+n_2-1} \sum_{j=0}^k   \binom{2n_1+2n_2-2k}{2n_1-2j-1}
 \binom{k}{j}^2 .
\een
By using TRR in genus zero, string equation and the divisor equation
one can show that \cite{Dij-Wit, Nor-Sco}:
\be \label{eqn:P1-two-point-g=0}
\begin{split}
& \corr{\tau_{2n_1}(Q)\tau_{2n_2}(Q)}_{0,2;n_1+n_2+1}
= \frac{q^{n_1+n_2+1}}{n_1+n_2+1}\frac{1}{n_1!n_2!}, \\
& \corr{\tau_{2n_1-1}(Q)\tau_{2n_2-1}(Q)}_{0,2;n_1+n_2}
= \frac{q^{n_1+n_2}}{n_1+n_2} \frac{n_1n_2}{n_1!n_2!}.
\end{split}
\ee
So one gets two combinatorial identities:
\ben
&& \frac{1}{(n_1+n_2+1)!^2}
\sum_{k=0}^{n_1+n_2} \sum_{j=0}^k
\binom{2n_1+2n_2-2k}{2n_1-2j} \binom{k}{j}^2
=  \frac{1}{n_1+n_2+1}\frac{1}{n_1!n_2!}, \\
&&  \frac{1}{(n_1+n_2)!^2}
\sum_{k=0}^{n_1+n_2-1} \sum_{j=0}^k
\binom{2n_1+2n_2-2k}{2n_1-2j-1} \binom{k}{j}^2
= \frac{1}{n_1+n_2} \frac{n_1n_2}{n_1!n_2!}.
\een
Reversely,
if one proves these identities,
then one can use them to get \eqref{eqn:P1-two-point-g=0}.

If we consider the terms with $\lambda^2$,
we get:
\ben
&& \sum_{d=0}^\infty q^d
\sum_{k_1,k_2 \geq 0} \corr{\tau_{k_1}(Q)\tau_{k_2}(Q)}_{1,2;d} z_1^{k_1+1}z_2^{k_2+1} \\
& = & -\frac{1}{24} \sum_{d=0}^\infty q^d   \frac{z_1^3z_2+z_1z_2^3}{d!^2}
\sum_{k=0}^{d-1} (z_1+z_2)^{2d-2-2k}
\sum_{j=0}^k \binom{k}{j}^2 z_1^{2j} z_2^{2k-2j} \\
& - & \frac{1}{24} \sum_{d=0}^\infty q^d  \frac{z_1z_2}{d!^2}
\sum_{k=0}^{d-1} (2d-2-2k) (z_1+z_2)^{2d-2k}
\sum_{j=0}^k \binom{k}{j}^2 z_1^{2j}z_2^{2k-2j}\\
& - & \frac{1}{24}\sum_{d=0}^\infty q^d  \frac{z_1z_2}{d!^2}
\sum_{k=0}^{d-1} (z_1+z_2)^{2d-2-2k}
\sum_{j=0}^k \binom{k}{j}^2 2jz_1^{2j+2} z_2^{2k-2j} \\
 & - & \frac{1}{24} \sum_{d=0}^\infty q^d  \frac{z_1z_2}{d!^2}
\sum_{k=0}^{d-1} (z_1+z_2)^{2d-2-2k}
\sum_{j=0}^k \binom{k}{j}^2z_1^{2j}(2k-2j)z_2^{2k-2j}.
\een
It becomes more involved  to extract information about the two-point correlators in higher genera.

Another way to extract some information from \eqref{eqn:P1-Two-Point} is to take
the coefficients of $z_1^k$ on both sides.
For $k=1$ we get:
\ben
&& \sum_{d=0}^\infty q^d \sum_{g=0}^\infty \lambda^{2g}
\sum_{k_2 \geq 0} \corr{\tau_{0}(Q)\tau_{k_2}(Q)}_{g,2;d} z_2^{k_2+1} \\
& = & \sum_{d=1}^\infty q^d  \frac{1}{d!(d-1)!\lambda^{2d-1}}
\zeta(\lambda z_2)^{2d-1}.
\een
The right-hand side is essentially Bessel function
of order one.
One can expand the right-hand side as follows:
\ben
&& \sum_{d=1}^\infty q^d  \frac{1}{d!(d-1)!\lambda^{2d-1}}
(e^{\lambda z_2/2} -e^{-\lambda z_2/2})^{2d-1} \\
& = & \sum_{d=1}^\infty q^d  \frac{1}{d!(d-1)!\lambda^{2d-1}}
\sum_{j=0}^{d-1} (-1)^j \binom{2d-1}{j}
(e^{(2d-1-2j)\lambda z_2/2}
- e^{-(2d-1-2j)\lambda z_2/2} )\\
& = & \sum_{d=1}^\infty q^d  \frac{1}{d!(d-1)!\lambda^{2d-1}}
\sum_{j=0}^{d-1} (-1)^j \binom{2d-1}{j}
\sum_{l =0}^\infty\frac{(2d-1-2j)^{2l+1}\lambda^{2l+1} z_2^{2l+1}}
{4^{l}(2l+1)!}
\een

Similarly,
the coefficients of $z^2$ on both sides of \eqref{eqn:P1-Two-Point}
give us
\ben
&& \sum_{d=0}^\infty q^d \sum_{g=0}^\infty \lambda^{2g}
\sum_{k_2 \geq 0} \corr{\tau_{1}(Q)\tau_{k_2}(Q)}_{g,2;d} z_2^{k_2+1} \\
& = & \sum_{d=0}^\infty q^d
\frac{\lambda \zeta(\lambda z_2) }{d!^2\lambda^{2d}}
\sum_{k=0}^{d-1} (2d-2-2k) \lambda
\zeta(\lambda z_2)^{2d-1-2k}  .
\een
One can get similar formulas for other $k=k_1+1$
in the same fashion.

\subsection{The formula of Dubrovin and Yang}

The examples in the last two Subsections
indicate that it involves some complicated combinatorial identities
to simplify the formulas derived from the formula of Okounkov and
Pandharipande.
Dubrovin and Yang define
\be
C_n(x_1, \dots, x_n; \lambda):
= \lambda^n \sum_{k_1, \dots, k_n \geq 0}
\corr{\tau_{k_1}(Q) \cdots \tau_{k_n}(Q)}
\prod_{i=1}^n \frac{(k_i+1)!}{x_i^{k_i+2}}.
\ee
They conjecture the following formulas.
Define a $2 \times 2$ matrix-valued series by
\be
R(x; \lambda) := \begin{pmatrix}
1 & 0 \\ 0 & 0 \end{pmatrix}
 + \begin{pmatrix}
\alpha(x;\lambda) & \beta(x;\lambda) \\
\gamma(x;\lambda) & -\alpha(x;\lambda)
\end{pmatrix}
\ee
where
\ben
\alpha(x; \lambda) & = &
\sum_{j=0}^\infty\frac{1}{4^j x^{2j+2}}
\sum^j_{i=0}   \frac{\epsilon^{2(j-i)}}{i!(i + 1)!}
\sum^i_{l=0}
(-1)^l(2i + 1 - 2l)^{2j+1}\binom{2i + 1}{l}, \\
\gamma(x; \lambda) & = &Q(x; \lambda) + P(x; \lambda), \\
\beta(x; \lambda)
& = & Q(x; \lambda) - P(x; \lambda),  \\
P(x; \lambda) &:=&
\sum_{j=0}^\infty
\frac{1}{4^j \lambda{2j+1}}
\sum^j_{i=0} \frac{\epsilon^{2(j-i)}}{i!^2}
\sum^i_{l=0}
(-1)^l(2i + 1 -2l)^{2j}
\biggl(\binom{2i}{l} -\binom{2i}{l- 1}\biggr), \\
Q(x; \epsilon) & := & -\frac{1}{2}
\sum_{j=0}^\infty
\frac{1}{4^j \lambda^{2j+2}}
\sum^j_{i=0}
\epsilon^{2(j-i)+1} \frac{2i + 1}{i!^2} \\
&& \cdot \sum^i_{l=0}
(-1)^l(2i + 1 - 2l)^{2j}
\biggl(\binom{2i}{l} - \binom{2i}{l- 1} \biggr).
\een
Then
\ben
&&C_2(x_1, x_2; \lambda) =
\frac{\tr [R(x_1; \lambda) R(x_2; \lambda)] - 1}
{(x_1 - x_2)^2},
\een
and for $n\geq 3$,
\ben
&& C_n(x_1, \dots , x_n; \lambda) = - \frac{1}{n}
\sum_{\sigma \in S_n}
\frac{\tr [R(x_{\sigma_1}; \lambda) \cdots
R(x_{\sigma_n}; \lambda)]}
{(x_{\sigma_1} - x_{\sigma_2})
\cdots (x_{\sigma_{n-1}} - x_{\sigma_n})
(x_{\sigma_n} - x_{\sigma_1})}.
\een
This conjecture has been proved
by Marchal \cite{Mar} using Eynard-Orantin topological recursions
and by Dubrovin-Yang-Zagier \cite{Dub-Yan-Zag} using Toda equations.

\subsection{Eynard-Orantin topological recursions
for stationary GW invariants of $\bP^1$}

Let us now recall some details about the Eynard-Orantin
topological recursions \cite{Eyn-Ora}
satisfied by the stationary Gromov-Witten invariants
of $\bP^1$, conjectured by Norbury and Scott \cite{Nor-Sco},
proved in \cite{D-B-O-S-S} and extended to the equivariant case
by Fang-Liu-Zong \cite{Fan-Liu-Zon}.
We will interpret some aspects of these results
from the point of view of emergent geometry.
In genus zero,
\be
\corr{\tau_{2n}(Q)}_0 = \frac{1}{(n+1)!^2},
\ee
we have seen in \S \ref{sec:OP-One-Point} that
\ben
W_{0,1}(x):=\sum_{n=0}^\infty \corr{\tau_{2n}(\omega)}_0 \frac{(2n+1)!}{x^{2n+2}}
& = & \sum_{n=0}^\infty \frac{(2n+1)!q^{2n+1}}{(n+1)!^2} \frac{1}{w^{2n+2}} \\
& = & \ln x- \ln \frac{x+\sqrt{x^2-4q}}{2}.
\een
Define
\be
y(x): = W_{0,1}(x) - \ln x -\ln q.
\ee
Then we have
\be
y = \ln \frac{x-\sqrt{x^2-4q}}{2},
\ee
and so
\be
e^y = \frac{x-\sqrt{x^2-4q}}{2},
\ee
and
\be
e^{-y} = \frac{x+\sqrt{x^2-4q}}{2q}.
\ee
Therefore,
\be
x = e^y + q e^{-y}.
\ee
If we set $z = e^y$,
then
\begin{align}
x & = z + \frac{q}{z}, & y & =\ln z.
\end{align}
This shows that the spectral curve is given by the one-point function in genus zero.
Next we show that Bergman kernel is given by the two-point function in genus zero.
By \eqref{eqn:P1-two-point-g=0}, we have
\ben
W_{0,2}(x_1,x_2;q)
& = & \sum_{n_1,n_2=0}^\infty \corr{\tau_{2n_1}(Q)\tau_{2n_2}(Q)}_0
\frac{(2n_1+1)!}{x_1^{2n_1+2}}\frac{(2n_2+1)!}{x_2^{2n_2+2}} \\
& + & \sum_{n_1,n_2=1}^\infty \corr{\tau_{2n_1-1}(Q)\tau_{2n_2-1}(Q)}_0
\frac{(2n_1)!}{x_1^{2n_1+1}}\frac{(2n_2)!}{x_2^{2n_2+1}} \\
& = & \sum_{n_1,n_2=0}^\infty  \frac{1}{n_1!^2n_2!^2}\frac{q^{n_1+n_2+1}}{n_1+n_2+1}
\frac{(2n_1+1)!}{x_1^{2n_1+2}}\frac{(2n_2+1)!}{x_2^{2n_2+2}} \\
& + & \sum_{n_1,n_2=1}^\infty  \frac{n_1n_2}{n_1!^2n_2!^2}\frac{q^{n_1+n_2}}{n_1+n_2}
\frac{(2n_1)!}{x_1^{2n_1+1}}\frac{(2n_2)!}{x_2^{2n_2+1}}.
\een
To take the summations,
we take derivatives with respect to $q$ to get:
\ben
\frac{\pd}{\pd q} W_{0,2}(x_1,x_2)
& = & \sum_{n_1,n_2=0}^\infty  \frac{1}{n_1!^2n_2!^2} q^{n_1+n_2}
\frac{(2n_1+1)!}{x_1^{2n_1+2}}\frac{(2n_2+1)!}{x_2^{2n_2+2}} \\
& + & \sum_{n_1,n_2=1}^\infty  \frac{n_1n_2}{n_1!^2n_2!^2}q^{n_1+n_2-1}
\frac{(2n_1)!}{x_1^{2n_1+1}}\frac{(2n_2)!}{x_2^{2n_2+1}} \\
& = & \sum_{n_1=0}^\infty  \frac{q^{n_1}}{n_1!^2}
\frac{(2n_1+1)!}{x_1^{2n_1+2}}\cdot
\sum_{n_2=0}^\infty  \frac{q^{n_2}}{n_2!^2}   \frac{(2n_2+1)!}{x_2^{2n_2+2}} \\
& + & q \sum_{n_1=1}^\infty  \frac{n_1}{n_1!^2}q^{n_1-1}
\frac{(2n_1)!}{x_1^{2n_1+1}}
\sum_{n_2=1}^\infty  \frac{n_2}{n_2!^2}q^{n_2-1}\frac{(2n_2)!}{x_2^{2n_2+1}} \\
& = & \frac{x_1}{(x_1^2-4q)^{3/2}} \cdot
\frac{x_2}{(x_2^2-4q)^{3/2}}\\
& +& q \cdot \frac{2}{(x_1^2-4q)^{3/2}}\cdot \frac{2}{(x_2^2-4q)^{3/2}}.
\een
After integration we get:
\be \label{eqn:W(0,2)}
W_{0,2}(x_1,x_2)
= \frac{x_1x_2-4q}{2\sqrt{x_1^2-4q}\sqrt{x_2^2-4q}(x_1-x_2)^2}
- \frac{1}{2(x_1-x_2)^2},
\ee
where the right-hand side of this equality is understood as a power series in $q$:
\ben
\frac{1}{2(x_1-x_2)^2}\biggl( (1-4q/(x_1x_2))(1-4q/x_1^2)^{-1/2}(1-4q/x_2^2)^{1/2}-1\biggr).
\een
The Bergman kernel is then given by:
\be
B(x_1,x_2) = \frac{dx_1dx_2}{(x_1-x_2)^2}
+ W_{0,2}(x_1,x_2)dx_1dx_2.
\ee
By straightforward computations we then get the following:

\begin{prop}
When $x_1=z_1 + q/z_1$ and $x_2 = z_2 + q/x_2$,
the Bergman kernel is given by
\be
B(x_1,x_2) = \frac{dz_1dz_2}{(z_1-z_2)^2}.
\ee
\end{prop}

So now all the needed data
\begin{align}
x(z) & = z + \frac{q}{z}, & y(z) & = \ln z, & B(z_1,z_2) & = \frac{dz_1dz_2}{(z_1-z_2)^2}
\end{align}
emerge from the computations of the one-point function and the two-point function
of stationary GW invariants of $\bP^1$.
As conjectured in \cite{Nor-Sco} and proved in \cite{D-B-O-S-S},
the Eynard-Orantin topological recursion  for such data produce
a sequences of multilinear differentials
$\omega_{g,n}(z_1, \dots, z_n)$.
After the change of coordinates $x_i =z_i+q/z_i$,
one has
\be
\omega_{g,n}(z_1, \dots, z_n)
= \sum_{k_1, \dots, k_n\geq 0} \corr{\tau_{k_1}(Q) \cdots \tau_{k_n}(Q)}
\prod_{i=1}^n \frac{(k_i+1)!dx_i}{x_i^{k_i+2}}.
\ee
See also \cite{Fan-Liu-Zon} and \cite{Bor-Nor}.

\subsection{Computations of general $n$-point functions of $\bP^1$}

Using Theorem \ref{thm:Main} one can compute the $n$-point function
of GW invariants of $\bP^1$:
They can all be reduced to the $n$-point function of
stationary GW invariants of $\bP^1$.
For example,
\ben
W(u_0) & = & \sum_{n=0}^\infty \corr{\tau_n(P)} \cdot \frac{(n+1)!}{u_0^{n+1}} \\
& = & \res_{x=0} (2L(u_0,x) W(x))
+ \half \res_{x=0}(2K(u_0,x)W(x,x)) \\
& + & \half \res_{x=0}(2K(u_0,x)W(x)W(x)) \\
& = &  \res_{x=0}(2 \log \big(1-\frac{x }{u_0}\big) \frac{du_0}{u_0-x}W(x)) \\
& + & \half \res_{x=0} \biggl(2 \frac{1}{u_0-x}\frac{du_0}{dx}W(x,x) \biggr) \\
& + & \half \res_{x=0} \biggl(2 \frac{1}{u_0-x}\frac{du_0}{dx}W(x)W(x) \biggr).
\een
For the genus zero part we have:
\ben
W_{0,1}(u_0) & = & \sum_{n=0}^\infty \corr{\tau_{2d-1}(P)}_{0,1;d}
\cdot \frac{(2d)!}{u_0^{2d}} \\
& = & \res_{x=0}\biggl(2 \log \big(1-\frac{x }{u_0}\big) \frac{du_0}{u_0-x}
\cdot W_{0,1}(x)  \biggr) \\
& + & \half \res_{x=0} \biggl(2 \frac{1}{u_0-x}\frac{du_0}{dx}W(x,x) \biggr) \\
& + & \half \res_{x=0} \biggl(2 \frac{1}{u_0-x}\frac{du_0}{dx}
\cdot  W_{0,1}(x) \cdot W_{0,1}(x) \biggr),
\een
where
\ben
W_{0,1}(x) = \ln\frac{x- \sqrt{x^2-4q}}{2} - \ln x
= \sum_{d=1}^\infty \frac{(2d-1)!}{d!^2} q^d x^{-2d} dx,
\een
and by \eqref{eqn:W(0,2)}, we have
\ben
W_{0,2}(x,x) = \frac{q}{(x^2-4q)^2}.
\een
A calculation gives us
\ben
W_{0,1}(u_0) & = & -2\sum_{d=1}^\infty \frac{(2d-1)!}{d!^2}\sum_{j=1}^{2d-1} \frac{1}{j}
u_0^{-2d}q^d \\
& + & \frac{q}{(u_0^2-4q)^2} + \biggl(\ln\frac{u_0- \sqrt{u_0^2-4q}}{2} - \ln u_0 \biggr)^2.
\een

\section{Concluding Remarks}

\label{sec:Conclusions}

In this note we have reformulated the strengthened
Virasoro constraints proved by Okounkov and Pandharipande
\cite{Oko-Pan-Vir}
as topological recursion relations
similar to the Eynard-Orantin topological recursions \cite{Eyn-Ora}.
In \cite{Oko-Pan-Vir} it is also shown that
there is another way to strengthen the standard
Virasoro constraints to the case of relative Gromov-Witten invariants
of the curves.
The reformulation can be done in exactly the same way.

In the standard Eynard-Orantin formalism
the insertions of odd observables do not seem to be have been considered in the literature.
So this work suggests an extension in this direction.

Another direction of research suggested by this work is to consider
the removal of the operators of the form $\tau_n(Q)$
in the correlators,
in other words,
to find topological recursions for
$$W(x, u_1, \dots, u_n, p_{i_1}^1, \dots, p_{i_k}^{k},\bar{p}_{j_1}^1, \dots, \bar{p}_{j_l}^l, x_1, \dots, x_m)$$
of the EO type. 

The earlier work \cite{Zhou} and this work have reformulated
the Virasoro constraints of the Gromov-Witten invariants
of a point and an algebraic curves respectively.
In a work in preparation we will generalize to other spaces
and Frobenius manifolds.

\vspace{.2in}
{\bf Acknowledgements}.
The author is partly supported by NSFC grants 11661131005
and 11890662.

\end{document}